\documentclass[aps,prl,reprint,floatfix,amsmath,amssymb,superscriptaddress]{revtex4-2}
\usepackage{amsmath,amssymb,latexsym,mathrsfs,amsfonts}
\usepackage{graphicx}
\usepackage{subfigure}

\usepackage{dcolumn}
\usepackage{epstopdf}
\usepackage{bm}
\usepackage{multirow}
\usepackage{color}
\usepackage{soul}
\usepackage{ulem}
\usepackage{mathtools}

\newcommand{\ssdw}{$s$SDW}

\begin{document}

\title{Coupled Spin-Density-Wave and Bond-Order Driven Metal-Insulator Transition in Altermagnetic CsCr$_2$S$_2$O}

\author{Chenchao Xu}
 \email[E-mail address: ]{chenchao-xu@hznu.edu.cn}
 \affiliation{School of Physics, Hangzhou Normal University, Hangzhou 310036, P. R. China}
 \affiliation{Center for Correlated Matter, Zhejiang University, Hangzhou 310058, China}

\author{Wansheng Bai}
 \affiliation{School of Physics, Hangzhou Normal University, Hangzhou 310036, P. R. China}

\author{Guo-Xiang Zhi}
 \affiliation{School of Physics, Hangzhou Normal University, Hangzhou 310036, P. R. China}
 \affiliation{Center for Correlated Matter, Zhejiang University, Hangzhou 310058, China}

\author{Yi Liu}
\affiliation{School of Physics, Zhejiang University, Hangzhou 310058, China}
\affiliation{Hangzhou International Innovation Institute, Beihang University, Hangzhou 311115, China}

\author{Xiaoqun Wang}
 \affiliation{School of Physics, Zhejiang University, Hangzhou 310058, China}
 \affiliation{Institute for Advanced Study in Physics, Zhejiang University, Hangzhou 310058, China}
 
\author{Jianhui Dai}
 \affiliation{School of Physics, Hangzhou Normal University, Hangzhou 310036, P. R. China}
 \affiliation{Institute for Advanced Study in Physics, Zhejiang University, Hangzhou 310058, China}

\author{Chao Cao}
 \email[E-mail address: ]{ccao@zju.edu.cn}
 \affiliation{Center for Correlated Matter, Zhejiang University, Hangzhou 310058, China}
 \affiliation{School of Physics, Zhejiang University, Hangzhou 310058, China}
 \affiliation{Institute for Advanced Study in Physics, Zhejiang University, Hangzhou 310058, China}
 
\date{\today}

\begin{abstract}
A metal-insulator transition (MIT) driven by bond order (BO) coupled with a {\it secondary} spin-density wave (SDW) is identified in CsCr$_2$S$_2$O. Such coupling is enabled as a result of the broken time-reversal symmetry due to the pre-existing C-type antiferromagnetic (C-AFM) order. First-principles calculations reveal an orbital-selective physics that Cr-$d_{yz}$ orbitals form local moments and establish the altermagnetic order, while the Cr-$d_{xz}$ orbitals remain metallic and hybridize with S-$p_z$. Thus the low-energy physics is governed by the Cr-$d_{xz}$ and S-$p_z$ orbitals. On-site interactions then enhance a {\it secondary} SDW (\ssdw) instability of the itinerant \(d_{xz}\) electrons, which couples to the Cr-$d_{xz}$--S-$p_z$ bonding order. The resulting coupled \ssdw-BO simultaneously produces experimentally observed structural distortion, charge disproportionation, local Cr-moment modulation, and gap opening. Our results establish an orbital-selective mechanism upon which pre-existing altermagnetism and electronic correlations cooperate to drive a structural MIT.
\end{abstract}

\maketitle

Metal-insulator transitions (MITs) are commonly associated with emergent long-range order, including charge-density-wave (CDW), magnetic ordering or SDW, bond or orbital ordering, and correlation-driven Mott transition~\cite{Imada1998,RevModPhys.40.677,RevModPhys.40.714,Kim2018MIT}. The most revealing cases, however, arise when these degrees of freedom coexist and interact~\cite{Biermann2005, McWhan1973, Mercy2017, Fagot2005, Radaelli2002, Rogge2018, Gorelov2010, Asamitsu1995, PhysRevLett.92.056402, Bansal2020}. In most cases, when magnetic ordering or SDW coexists with structural distortion or charge order, the magnetic ordering or SDW occurs at lower temperature after the structural distortion or charge order is formed, due to dominant energy scale of the latter\cite{Mercy2017,Asamitsu1995,PhysRevB.58.847,PhysRevB.62.844,PhysRevB.87.214410,PhysRevB.62.5619}. In other cases, structural distortion or charge order occurs almost simultaneously with magnetic ordering or SDW, as a result of reduced symmetry~\cite{Huang_BaFe2As2,Goldman_CaFe2As2,Jesche_SrFe2As2,Warmuth2018, Bansal2020,Liu:2024aa}. It is, however, quite rare to observe a pre-existing magnetic order to activate a subsequent charge-density-wave or structural transition.


The recently synthesized layered chromium oxychalcogenides $A$Cr$_2X_2$O ($A$ = Rb, Cs; $X$ = S, Se) provide a platform to address this question\cite{Liu2026CsCr2S2O,Sun2026RbCr2Se2O}. These materials are isostructural to $A$V$_2X_2$O, which has attracted considerable interest for its intertwined magnetic and electronic states, in particular the possible hidden $d$-wave altermagnetism due to inter-layer coupling\cite{Doan, Frandsen2014, YajimaBi2013, Ablimit2018RbV2Te2O, jiangMetallicRoomtemperatureDwave2025, zhangCrystalsymmetrypairedSpinValley2025}. In contrary, CsCr$_2$S$_2$O exhibits an unambiguous C-type antiferromagnetic (C-AFM) state and retains its altermagnetic character across a near-room-temperature metal-to-insulator transition\cite{Liu2026CsCr2S2O}. Magnetic order develops first at $T_{\mathrm N}=326$~K, followed by an intriguing structural distortion and MIT at $T_{\mathrm{MI}}=305$~K. The low temperature phase breaks the $C_4$ symmetry, and modulates the magnitude of the Cr-local moments. Despite of the close proximity between $T_{\mathrm{N}}$ and $T_{\mathrm{MI}}$, the relationship between the MIT and the C-AFM order remains an open question. In addition, previous calculations have shown that both on-site Coulomb interactions and sulfur displacement are required to drive the MIT, but the microscopic mechanism leads to such MIT is still unclear. Furthermore, why such MIT is only present in Cr-based compounds but absent in the V-based isostructural $A$V$_2X_2$O?

In this {\it Letter}, we show that neither is sulfur-displacement a property of the nonmagnetic lattice, nor is the MIT conventional purely correlation driven. Instead, the orbital selectivity assigns distinct roles to the two low-energy Cr $d$-orbitals.  The localized $d_{yz}$ orbitals predominantly support the pre-existing C-AFM background, whereas the metallic $d_{xz}$ orbitals hybridize with S-$p_z$ and develop \ssdw\ fluctuations as correlations increase. Because the \ssdw\ fluctuations reside in the Cr--S-oriented $d_{xz}$ orbitals, they couple to the Cr-$d_{xz}$--S-$p_z$ BO and ultimately to the lattice. Charge disproportionation, Cr-moment modulation, and gap opening thus emerge as different facets of this coupled instability, rather than as three independent transitions. This mechanism provides a microscopic connection between C-AFM and the near-room-temperature structural distortion as well as the MIT. In addition, it naturally explains why such transition is absent in V-based compounds.

Our calculations were performed within density functional theory (DFT) using the \textsc{Vienna Ab initio Simulation Package} (VASP)\cite{VASP_Kresse_PRB93,VASP_Kresse_PRB99}.  A plane-wave energy cutoff of 600~eV and a $\Gamma$-centered $12\times12\times6$ $\mathbf{k}$-point mesh were used for the high-symmetry ($P4/mmm$) structure, together with the projector-augmented-wave (PAW) method~\cite{PAW}.  The atomic coordinates were fully relaxed until the force on each atom was below 1~meV/\AA\ and the residual stress was below 0.1~kbar.  The DFT bands derived from the Cr-$3d$, S-$3p$, and O-$3p$ orbitals were fitted to a tight-binding (TB) Hamiltonian using the maximally projected Wannier-function method~\cite{Wannier90}.  The unfolded band structures and multiorbital susceptibilities were calculated from the symmetrized TB Hamiltonian~\cite{ZHI2022108196}.  Finite-temperature phonons were treated within the stochastic self-consistent harmonic approximation (SSCHA)~\cite{Monacelli2021SSCHA, Errea2014SSCHA, Bianco2017SSCHA, Monacelli2018SSCHA}, using dynamical matrices obtained from density functional perturbation theory (DFPT)~\cite{Giustino2017DFPT} on a $2\times2\times2$ $\mathbf{q}$-point grid at $U=2$~eV, with the S atoms placed at the minimum of the potential-energy surface.

\begin{figure}[t]
  \centering
  \includegraphics[width=\columnwidth]{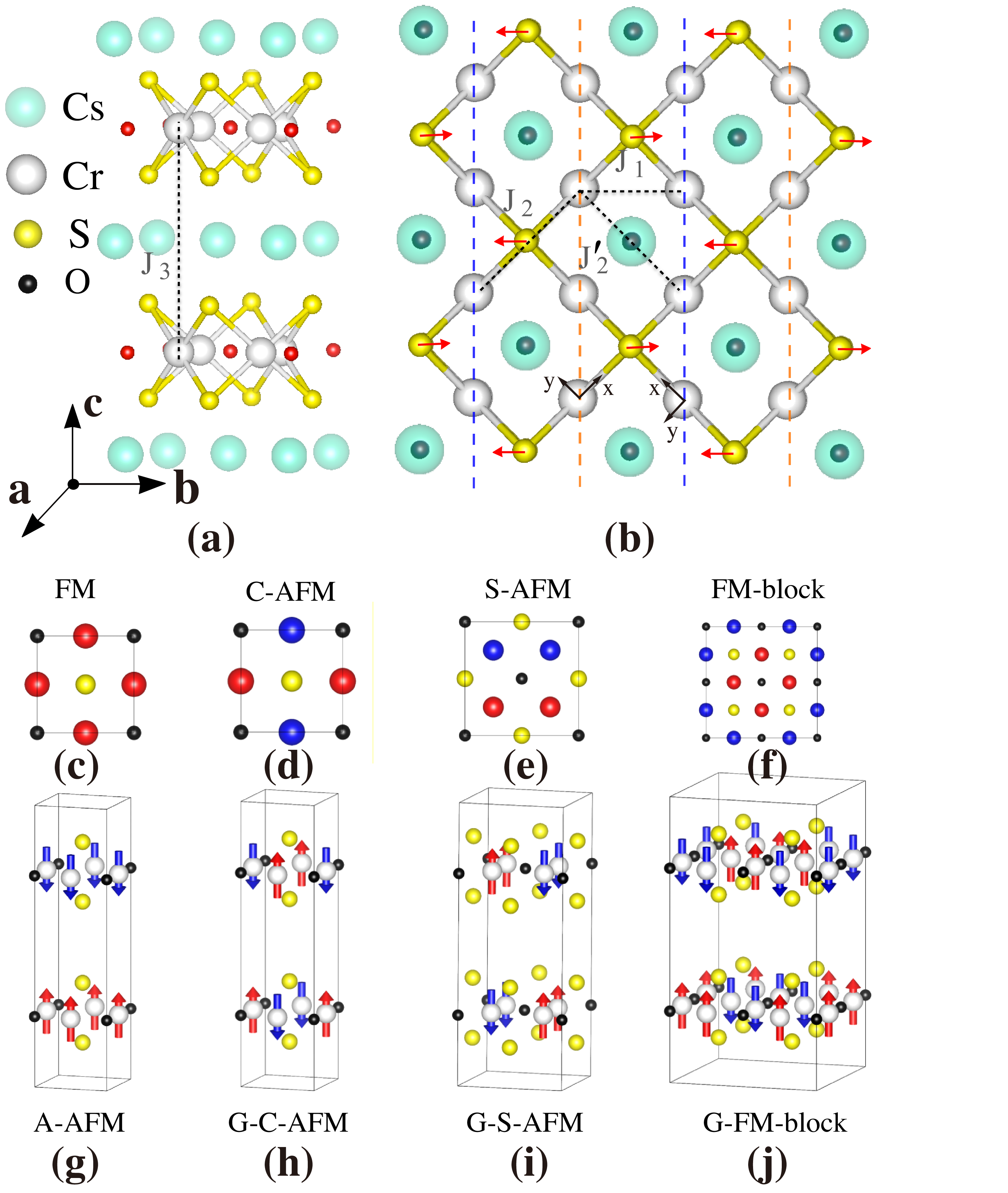}
  \caption{Crystal and magnetic structures of CsCr$_2$S$_2$O. (a) Side view of the layered crystal structure. (b) Top view of a [Cr$_2$S$_2$O] slab, showing the in-plane exchange pathways $J_1$, $J_2$, and $J_2'$; $J_3$ denotes the interlayer exchange. The black arrows indicate the local axes, with $x$ pointing from Cr toward S and $y$ from Cr toward O. The $d_{xz}$ and $d_{yz}$ orbitals are defined with respect to  local coordinate of each Cr atom. Red arrows denote the collective in-plane displacement of the two S atoms. The blue dashed lines label the Cr$_2$ columns that the S atoms approach, whereas the orange dashed lines label the Cr$_1$ columns from which they recede. (c-j) Typical collinear magnetic configurations considered in the total-energy calculations. Red and blue spheres (arrows) indicate opposite Cr-spin orientations.}
  \label{fig:structure_magnetism}
\end{figure}

CsCr$_2$S$_2$O crystallizes in the tetragonal CeCr$_2$Si$_2$C-type (1221) structure with space group $P4/mmm$ (No.~123), in which [Cr$_2$S$_2$O]$^{-}$ slabs alternate with Cs layers along the $c$ axis.  Within each slab, the Cr$_2$O plane adopts an anti-Cu$_2$O geometry, with S atoms capping the Cr squares from above and below.  Each Cr is octahedrally coordinated by 4 apical S and 2 in-plane O atoms, and the linear O--Cr--O axes are orthogonal on the two Cr sublattices [Fig.~\ref{fig:structure_magnetism}(a,b)]. To facilitate the following discussions, we define local-axis for Cr atoms such that the local $y$-axis is along Cr-O bonds and the local $z$-axis is along $c$-axis. Neutron diffraction experiments established C-type antiferromagnetic order below $T_{\mathrm N}=326$~K \cite{Liu2026CsCr2S2O}. To examine its energetic stability, we calculate representative collinear spin configurations [Fig.~\ref{fig:structure_magnetism}(c--j)] at $U=0$, 2, and 4~eV. Their energies are mapped onto an effective Heisenberg model (see Supplementary Information for detail)
 $$\mathcal{H}=J\sum_{\langle i,j\rangle}\mathbf{S}_i\cdot\mathbf{S}_j,$$
 where $\langle i,j\rangle$ denotes the nearest neighboring sites $i$ and $j$.  C-AFM is the lowest-energy state throughout $U\in [0,4]$ eV.  The dominant in-plane nearest-neighbor coupling is antiferromagnetic, with $J_1=25.3$, 13.2, and 5.2~meV/$S^2$ for $U=0$, 2, and 4~eV, respectively, whereas the interlayer coupling remains weakly ferromagnetic, $J_3\simeq-0.3$ to $-0.4$~meV/$S^2$. Other longer-range exchange couplings increase as $U$ grows, indicating that correlations soften the magnetic energy landscape even though the C-AFM ground state remains robust.

\begin{figure}[t]
  \centering
  \includegraphics[width=\columnwidth]{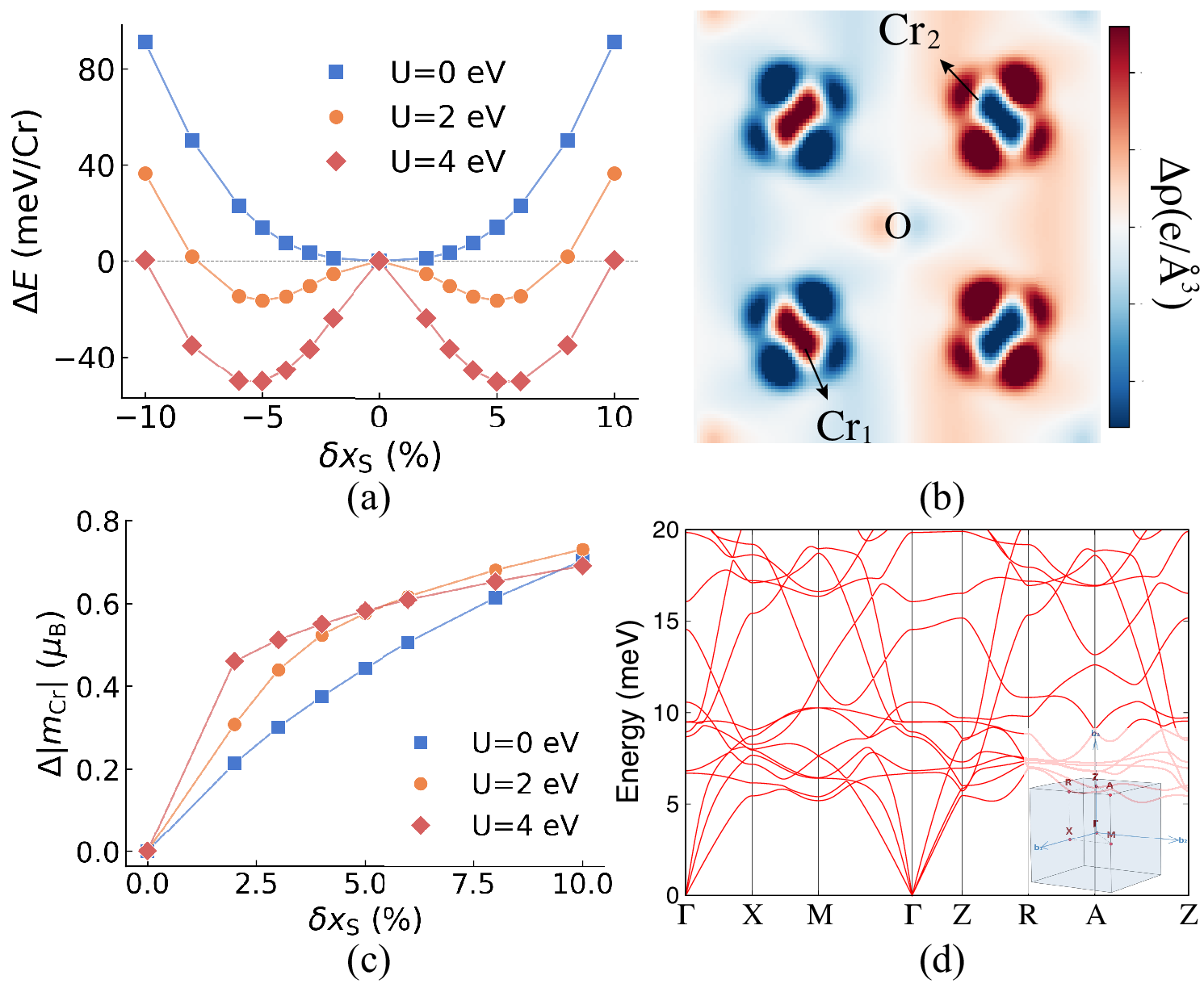}
  \caption{Correlation-induced structural, charge, and magnetic instabilities in C-AFM CsCr$_2$S$_2$O. (a) Total-energy variation with the displacement $\delta x_{\mathrm S}$ of the two S atoms along the direction indicated by the red arrows in Fig.~\ref{fig:structure_magnetism}(b), calculated at $U=0$, 2, and 4~eV. $\delta x_{\mathrm S}=\Delta x_{\mathrm S}/d_{\mathrm{Cr-Cr}}\times100\%$, where $\Delta x_{\mathrm S}$ is the displacement of each S atom and $d_{\mathrm{Cr-Cr}}$ is the in-plane nearest-neighbor Cr-Cr distance in the undistorted structure. Double-well energy profiles develop for $U=2$ and 4~eV, with minima near $|\delta x_{\mathrm S}|=5\%$. (b) Charge-density difference between $\delta x_{\mathrm S}=0\%$ and 5\% at $U=2$~eV; red and blue denote charge accumulation and depletion, respectively. Cr$_1$ and Cr$_2$ are marked by the orange and blue dashed lines, respectively, in Fig.~\ref{fig:structure_magnetism}(b). (c) Difference between the local magnetic moments on Cr$_1$ and Cr$_2$ as a function of $\delta x_{\mathrm S}$. (d) Phonon spectrum of the structure with $\delta x_{\mathrm S}=5\%$ at $U=2$~eV and $T=150$~K.}
  \label{fig:structural_instability}
\end{figure}

The structural and electronic instability is examined within the nonmagnetic phase and the C-AFM phase by collectively displacing the pair of S atoms flanking the Cr$_2$O layer along an in-plane nearest-neighbor Cr–Cr bond direction [red arrows in Fig.~\ref{fig:structure_magnetism}(b)]. In the C-AFM state, the high-symmetry position of S becomes a local maximum and a double-well potential develops for $U=2$ and 4~eV [Fig.~\ref{fig:structural_instability}(a)].  This instability is consistent with the observed tetragonal-to-orthorhombic transition to the $Pmam$ phase observed by Liu, {\it et al}.\cite{Liu2026CsCr2S2O}. In contrast, the C-AFM calculation at $U=0$ and all nonmagnetic calculations retain a single-well potential (see Supplementary Information for detail). Therefore, the transition requires both a pre-existing magnetic background and sufficiently strong electronic correlations.  This hierarchy reflects the experimental sequence in which C-AFM order appears at $T_{\mathrm N}=326$~K before the structural distortion and MIT at $T_{\mathrm{MI}}=305$~K.

The structural distortion simultaneously reorganizes charge and spin. At $U=2$~eV and $\delta x_{\mathrm S}=5\%$, the initially equivalent Cr sites acquire a Bader-charge difference of approximately $0.15~e^-$ [Fig.~\ref{fig:structural_instability}(b)]~\cite{Sanville2007,Tang2009}. Their moments also become inequivalent, and the difference in magnitudes grows monotonically with displacement [Fig.~\ref{fig:structural_instability}(c)]. This inequivalence constitutes a \ssdw\ superposed on the primary C-type altermagnetic order. At the double-well minimum, the finite-temperature phonon spectrum at 150~K is free of imaginary modes [Fig.~\ref{fig:structural_instability}(d)], confirming the dynamical stability of this phase. Charge disproportionation, moment-magnitude modulation, and S displacement therefore constitute a composite ordered state of bond, charge and spin, instead of three independent transitions.

\begin{figure}[t]
  \centering
  \includegraphics[width=\columnwidth]{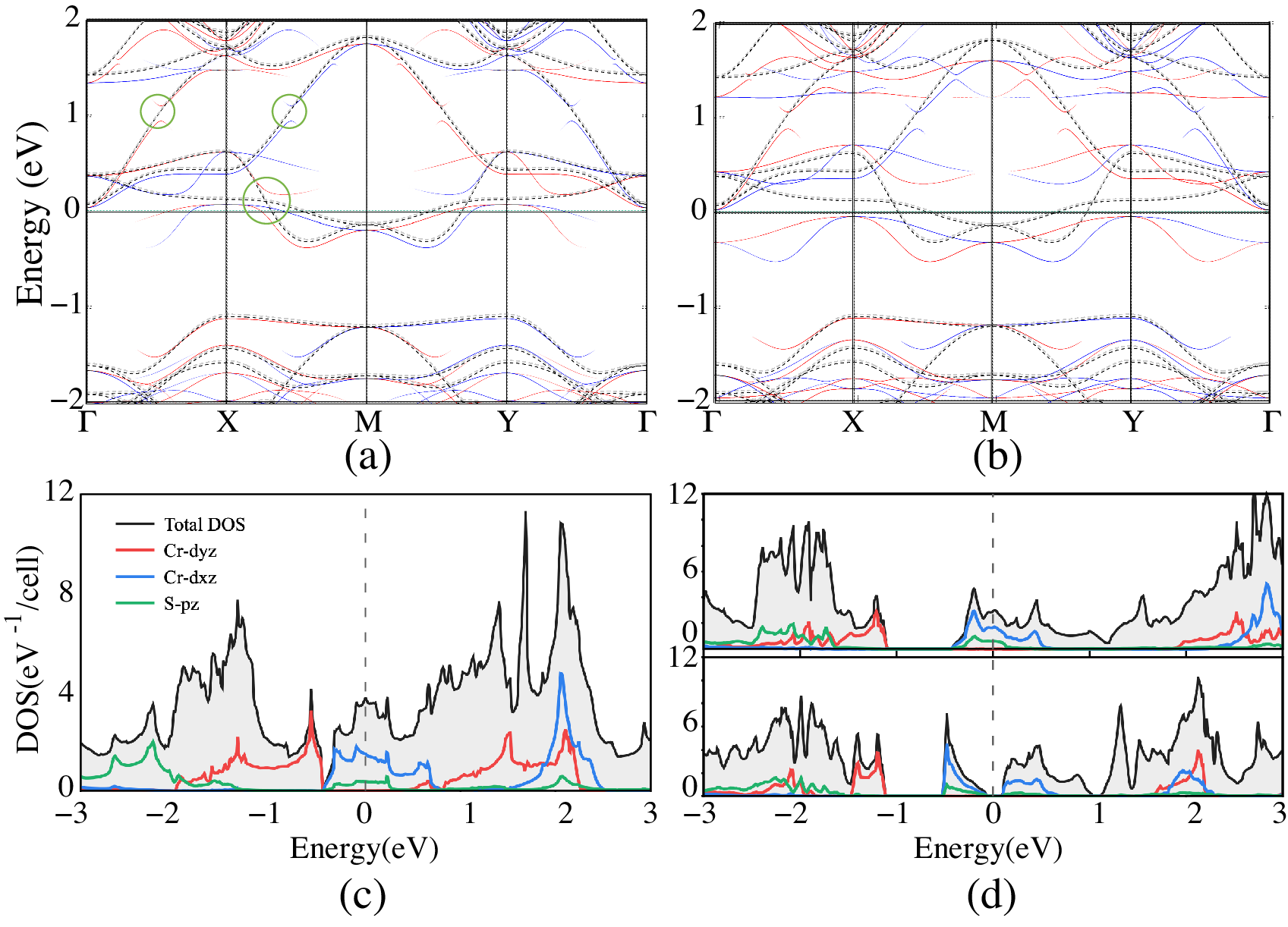}
  \caption{Electronic structures of CsCr$_2$S$_2$O unfolded to the $P4/mmm$ primitive cell and the corresponding orbital-resolved DOS. Gray dashed curves denote the bands of undistorted $P4/mmm$ state. (a) At $\delta x_{\mathrm S}=2\%$, $C_4$-symmetry breaking opens direct gaps at the band crossings marked by green circles at $U=2$~eV. (b) At the double-well minimum ($\delta x_{\mathrm S}=5\%$), a full gap opens  at $U=2$~eV. (c) DOS of the undistorted $P4/mmm$ state at $U=0$~eV, showing that the Cr-$d_{yz}$ states are split to either side of $E_{\mathrm F}$, while the finite DOS at $E_{\mathrm F}$ arises primarily from hybridized Cr-$d_{xz}$ and S-$p_z$ states. The Cr orbitals are labeled in the local axes defined in Fig.~\ref{fig:structure_magnetism}(b). (d) DOS corresponding to (a) and (b) in the upper and lower panels, respectively. The S-$p_z$ contributions in
(c,d) are multiplied by 2.  S is displaced strictly parallel to the Cr--O plane; full relaxation, which allows an out-of-plane S displacement toward the Cr--O layer, shifts the onset of the global gap to slightly larger $U$~\cite{Liu2026CsCr2S2O}.
}  
  \label{fig:band_dos}
\end{figure}

To determine how this coupled order reconstructs the electronic structure, we unfold the electronic band structure of the distorted supercell onto the $P4/mmm$ primitive Brillouin zone (Fig.~\ref{fig:band_dos}).  Although the S displacement breaks $C_4$, the altermagnetic spin splitting survives because it is protected by the gliding mirror symmetries of the $Pmam$ phase\cite{Liu2026CsCr2S2O}.  The loss of sublattice equivalence opens direct gap at the band crossings [Fig.~\ref{fig:band_dos}(a,b)]. The system remains metallic for small displacements and becomes insulating near the double-well minimum, $\delta x_{\mathrm S}=5\%$. The orbital-resolved DOS of the undistorted $U=0$ C-AFM state [Fig.~\ref{fig:band_dos}(c)] uncovers that the Cr-$d_{yz}$ states are local states, separated by a gap of 650 meV at the Fermi level, whereas the Cr-$d_{xz}$ and S-$p_z$ states remain metallic and hybridize within $E_F-$ 0.45 eV to $E_F+$ 0.8 eV. This separation is consistent with an orbital-selective picture: the localized $d_{yz}$ orbitals predominantly support the primary C-AFM order, while the itinerant $d_{xz}$--$p_z$ states form the active low-energy channel. As we shall show below, correlations enhance \ssdw\ fluctuations of $d_{xz}$ orbitals as well as the Cr-S BO at the same characteristic ordering vector $\mathbf{Q}=(\pi, \pi, 0)$ on the C-AFM background. In addition, the coupling between the \ssdw\ order and the BO is activated by the C-AFM background, leading to a MIT due to the coupled order.

We now characterize the electronic precursor of the C-AFM state through static susceptibility calculations.  The \ssdw\ susceptibility $\chi_{\mathrm{amp}}(\mathbf{q})$ (see Supplementary Information for details) peaks at $\mathbf{M}=(\pi,\pi,0)$ already at the bare level and is further enhanced by RPA correlations. This response implies a tendency toward an additional modulation of the Cr-moment amplitudes on the C-AFM background. To characterize the electronic bonding response associated with the sulfur displacement, we also introduce the local Cr-$d_{xz}$--S-$p_z$ bonding operator
$$\hat{\mathcal O}^{\mathrm{bond}}=\sum_\sigma(d_{i\sigma}^\dagger p_{j\sigma}+p_{j\sigma}^\dagger d_{i\sigma}),$$
where $d_i$ denotes the Cr $d_{xz}$-orbital at site $i$ and $p_j$ the S-$p_z$ orbital at site $j$ neighboring to $i$ (see Supplementary Information for detail).  The sulfur-displacement pattern selects one of the two in-plane diagonals and hence constitutes a $C_4$-breaking Cr--S BO.  The Cr--S BO susceptibility is calculated in a $\sqrt{2}\times\sqrt{2}$ supercell, for which the primitive-cell $M$ point folds to $\Gamma$.  Interactions also enhance this Cr--S bonding response at $\Gamma$.  In contrast, the corresponding nonmagnetic response of CsCr$_2$S$_2$O likewise shows no enhancement (see Supplementary Information for detail). Therefore, both the \ssdw\ and bonding susceptibilities are enhanced at the ordering wave vector of the structural distortion, in the same magnetic and interaction regime where the double-well instability develops, suggests an effective coupling between the two fluctuations.

This coupling can be described by a minimal phenomenological free energy
\begin{equation}
\begin{aligned}
F={}&F_0+\frac{r_{\mathrm{amp}}}{2}(\delta\mathcal M)^2
+\frac{r_{\mathrm{bond}}}{2}\Phi_{\mathrm{bond}}^2
+g\mathcal{M}_0\,\delta\mathcal M\,\Phi_{\mathrm{bond}}, \nonumber
\end{aligned}
\label{eq:spin_bond_landau}
\end{equation}
where $\mathcal{M}_0$ is the pre-existing C-AFM order, supported predominantly by the localized $d_{yz}$ orbitals, $\delta\mathcal M\equiv\langle\hat{\mathcal A}(\mathbf M)\rangle$ is the $M$-point \ssdw\ order parameter carried by the metallic $d_{xz}$ states, and $\Phi_{\mathrm{bond}}\equiv\langle\hat{\mathcal O}^{\mathrm{bond}}(\mathbf M)\rangle$ is the Cr--S BO parameter. The coupling term satisfies the relevant symmetry constraints. In momentum space, $\mathcal M_0$ is uniform ($\Gamma$) while $\delta\mathcal M$ and $\Phi_{\mathrm{bond}}$ both modulate at $M$; their product therefore conserves crystal momentum. In real space, the S displacement toward Cr$_2$ suppresses its moment while the retreat from Cr$_1$ enhances its moment, showing that the two modulations are locked in phase. Moreover, the coupling is time-reversal even because $\mathcal M_0$ and $\delta\mathcal M$ are time-reversal odd, whereas $\Phi_{\mathrm{bond}}$ is even. Thus, this coupling is present only in the magnetically ordered phase, where it linearly mixes the \ssdw\ and bonding fluctuations. Consistently, no double-well potential appears either in the nonmagnetic calculations ($\mathcal{M}_0=0$) or when the \ssdw\ modulation is suppressed by constraining $\delta\mathcal M=0$ (Supplementary Information). Minimizing the free energy with respect to $\delta\mathcal M$ gives
\begin{equation}
 \delta\mathcal M
 =-\frac{g\mathcal{M}_0}{r_{\mathrm{amp}}}\Phi_{\mathrm{bond}},
 \nonumber
\end{equation}
and hence renormalizes the bonding stiffness to
\begin{equation}
 r_{\mathrm{bond}}^{\mathrm{eff}}
 =r_{\mathrm{bond}}-\frac{g^2\mathcal{M}_0^2}{r_{\mathrm{amp}}}. \nonumber
 \label{eq:effective_bond_stiffness}
\end{equation}
Since $\chi_{\mathrm{amp}}(M)=r_{\mathrm{amp}}^{-1}$, enhanced \ssdw\
fluctuations lower
$r_{\mathrm{bond}}^{\mathrm{eff}}$ and increase
$\chi_{\mathrm{bond}}(M)=1/r_{\mathrm{bond}}^{\mathrm{eff}}$.
The coupled mode becomes unstable when $r_{\mathrm{bond}}r_{\mathrm{amp}}=g^2\mathcal{M}_0^2$; its proximity to criticality is therefore controlled jointly by the uncoupled bonding stiffness, the \ssdw\ response, the pre-existing C-AFM order, and their coupling.

\begin{figure}[t]
  \centering
  \includegraphics[width=\columnwidth]{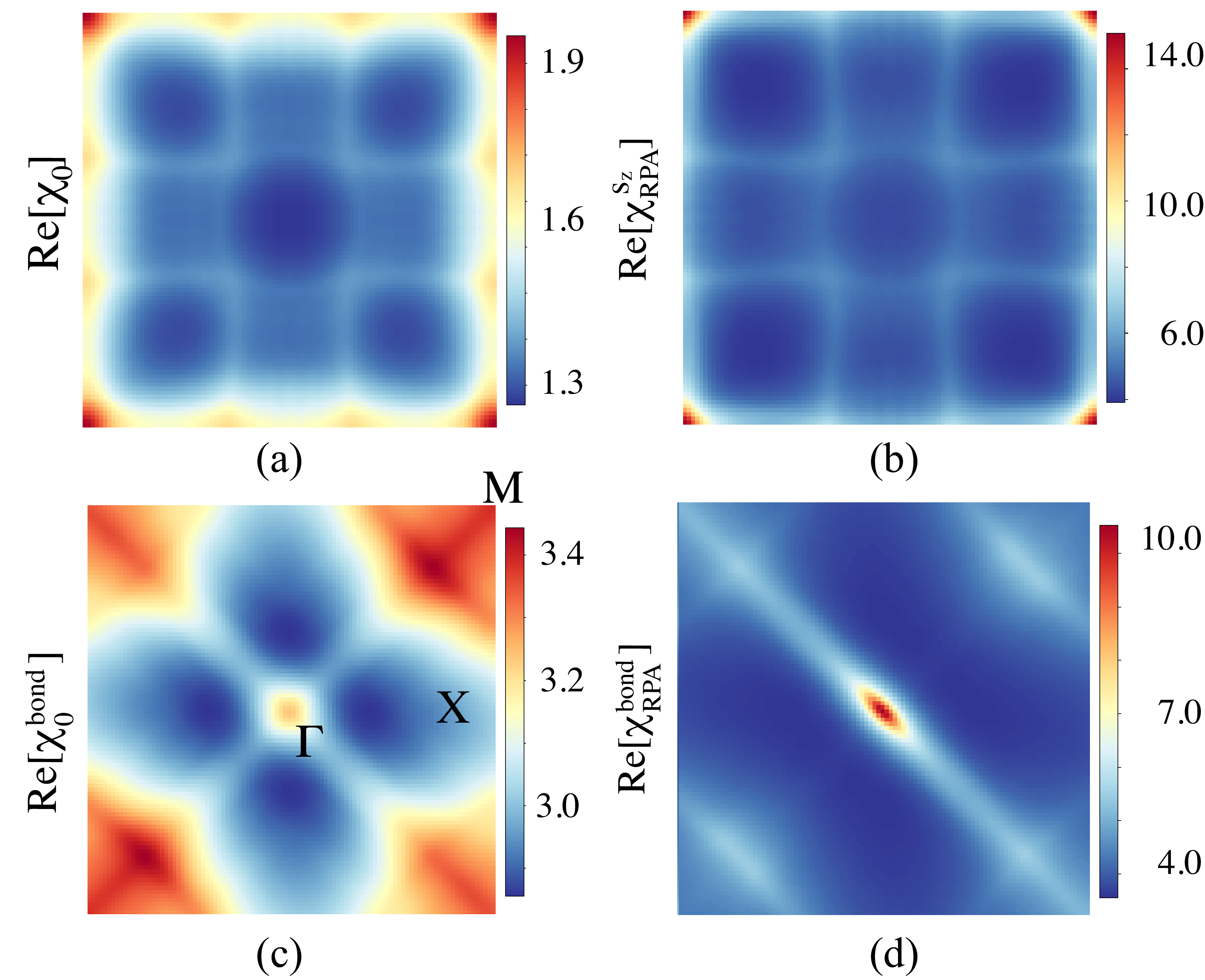}
  \caption{Static susceptibilities of C-AFM CsCr$_2$S$_2$O in the $k_z=0$
  plane. (a,b) Cr-$d_{xz/yz}$ \ssdw\ susceptibility
  $\chi_{\mathrm{amp}}(\mathbf q)$ calculated in the magnetic
  primitive cell: (a) bare and (b) RPA susceptibilities for $U=U'=3$~eV.
  (c,d) local Cr-$d_{xz}$--S-$p_z$ bonding susceptibility calculated in the
  $\sqrt{2}\times\sqrt{2}$ supercell: (c) bare and (d) RPA susceptibilities for
  $U=U'=3$~eV.  The supercell $\Gamma$ point in (c,d) corresponds to the
  primitive-cell $M=(\pi,\pi,0)$ point.}
  \label{fig:susceptibility}
\end{figure}

Our framework naturally accounts for the structural distortion is absent in KV$_2$Se$_2$O, although an analogous \ssdw\ instability at $M$-point is still present on its antiferromagnetic background\cite{jiangMetallicRoomtemperatureDwave2025, zhangCrystalsymmetrypairedSpinValley2025, r8nc-dpt8, llrq-1k9k, Ablimit2018RbV2Te2O, Wang2025AtomicSpinSensing, Fu2025AtomicScaleVisualization, Yang2026VisualizingSpinPolarization}. In KV$_2$Se$_2$O, the metallic states near $E_{\mathrm F}$ and their correlation-enhanced \ssdw\ fluctuations are dominated by V-$d_{yz}$ rather than the V-$d_{xz}$ orbitals (Supplementary Information). In addition, the V-$d_{xz}$ orbitals are completely unoccupied. Thus, although the \ssdw\ fluctuations are enhanced by on-site interactions in both compounds, only in CsCr$_2$S$_2$O does the fluctuating orbitals couple to the metal--ligand bonding channel, allowing the \ssdw\ fluctuation to induce the BO and the associated MIT.

In summary, CsCr$_2$S$_2$O hosts a coupled instability in which different Cr orbitals play distinct roles. The localized $d_{yz}$ orbitals predominantly support a pre-existing C-AFM background, whereas the metallic $d_{xz}$ states carry both the low-energy \ssdw\ fluctuation and the Cr--S BO. The Cr on-site interactions enhance both responses, while the background magnetism activates the coupling between them and transfer the resulting energy gain to the sulfur displacement. The ordered state consequently combines structural distortion, charge disproportionation, Cr-moment modulation, and gap opening while preserving altermagnetic spin splitting. This mechanism is consistent with the ordering sequence $T_{\mathrm N}>T_{\mathrm{MI}}$ and suggests that enhanced precursor fluctuations at $M$-point may develop between the two transitions, potentially manifested as phonon softening or diffuse scattering. Microscopically, orbital selectivity partition the pre-existing magnetic background and the critical fluctuation into different orbital sectors. Whether a \ssdw\ fluctuation can drive a bond--lattice transition is then controlled not only by its wave vector and strength, but also by the symmetry compatibility of the fluctuating orbital with the ligand--metal bonding channel.  This provides a microscopic criterion for coupled structural and metal--insulator transitions in multiorbital magnets.


The authors would like to thank Yu Song,  Jinke Bao, H.D. Wang and Jiangfan Wang for helpful discussions. All calculations were performed at the High Performance Computing Center and Advanced Computing Center of Hangzhou Normal University and 
High Performance Computing Center at the Center of Correlated Matters Zhejiang University. This work was supported by the National Key R\&D Program of China (Nos. 2024YFA1408303 \& 2022YFA1402202) and the National Natural Science Foundation of China (Nos. 12304175, 12274364, 12474132 \& 12004337).

\bibliographystyle{apsrev4-2}
\bibliography{1221}


\end{document}